\documentstyle{article}

\begin{document}
\pagestyle{myheadings}

\title{
Velocity Dominance Near a Crushing Singularity
\thanks{ Supported in part by NSF Grant PHY 9201196 to
Yale University.
To appear in the Lanczos Centenary Proceedings.}}
\author{Boro Grubi\v{s}i\'c \thanks{Yale University,
Department of Physics, 217 Prospect St., New Haven, CT} }
\date{ gr-qc/9404056 \\ \ \\ April 27, 1994}
\maketitle

\markboth{ Grubi\v{s}i\'c}{Velocity Dominance Near a
 Crushing Singularity}
\thispagestyle{empty}

\begin{abstract}
The asymptotic behavior of geometry near the boundary of maximal
Cauchy development is studied using a perturbative method, which at
the zeroth order reduces Einstein's equations to an exactly solvable
set of equations---Einstein's equations with all ``space" derivatives
dropped. The perturbative equations are solved to an {\em
arbitrarily-high order} for the cosmological spacetimes admitting
constant-mean-curvature foliation that ends in a crushing singularity,
i..e., whose mean curvature blows up somewhere.  Using a ``new" set of
dynamical variables (generalized Kasner variables)   restrictions on
the initial data  are found that make the zeroth-order term in the
expansion asymptotically dominant when approaching the crushing
singularity. The results obtained are in agreement with the first
order results of Belinskii, Lifshitz and Khalatnikov on general
velocity-dominated cosmological singularities, and, in addition,
provide clearer geometrical formulation.
\end{abstract}

\section{Introduction}

The set of all solutions of Einstein's equations is too large from
both the mathematical and physical standpoint. Many attempts have been
made to formulate the conditions that would restrict the solutions of
Einstein's equations to a class that is:

\begin{itemize}
\item{}large enough to include all ``reasonable"
physical situations,
\item{} small enough to exclude those deemed physically
``pathological",
\item{} mathematically as simple as possible.
\end{itemize}
 The most promising candidate to  date is (a generic subset of) the
class of maximally-extended  globally-hyperbolic spacetimes. 	The
most serious problem---the problem that initiated the Strong Cosmic
Censorship conjecture---for that class stems from the
Hawking-Penrose-Geroch singularity theorems, which indicate that the
globally-hyperbolic spacetimes, due to the possible incompleteness of
{\em non-curvature-singular} timelike geodesics, might have to be
extended beyond the boundary of their maximal Cauchy development.
Possible incompleteness of {\em curvature-singular} geodesics is
physically acceptable for two reasons. The first is that  being
crushed or torn apart by infinite tidal forces is believed to be a
good reason for any observer to disappear,  and the second is that
there is some hope of rescue by a quantum modification of classical
general relativity; as it is believed that quantum effects become
dominant in the regions of strong curvature, and might eliminate the
unpleasant classical infinite-curvature singularities.

It is, therefore, crucial to establish how often and under what
conditions the non-curvature-singular (bad), as opposed to the
curvature-singular (good), incomplete timelike geodesics occur  near
the boundary of the maximal Cauchy development; making the study of
the asymptotic behavior of geometry near the Cauchy boundary crucial
for the acceptability of the class of maximal globally-hyperbolic
spacetimes in general relativity. Unfortunately, due to the complexity
of Einstein's equations, an exact or useful description of  the
asymptotic behavior seems to be unattainable, at least so in the
forseeable future. Some results of  the work done by Belinski,
Khalatnikov and Lifshitz \cite{bkl82}, hereafter BKL, indicate that,
generically, the approach to the boundary of the maximal Cauchy
development (the singularity in their terminology) is extremely complex
and can be thought of as a sequence of pointwise Mixmaster-like
transitions between Kasner epochs. A satisfactory geometrical
formulation of their method is still lacking, and it  is still
controversial whether their method can give any information about the
global structure of the singularity \cite{barrow-tipler}.
Nevertheless, an important discovery of BKL was that the asymptotic
dynamics simplifies significantly for spacetimes for which, in a
suitable foliation, the spatial derivative terms in the equations of
motion can be  neglected asymptotically. Since  Einstein's equations
thus truncated are exactly solvable, we can extract all the asymptotic
properties from the solution of the truncated equations, the
Generalized Kasner Solution; a solution which evolves pointwise like
the Kasner solution. Most importantly, the Generalized Kasner solution
is generically curvature-singular, as there is only one, (1,0,0),
non-curvature-singular Kasner solution.

The question I shall try to answer here, without going into technical
details, is: Under what conditions is that approximation, usually
called the Velocity-Dominated Approximation (VDA) \cite{els72}, valid
in the class of globally-hyperbolic vacuum cosmological spacetimes,
when approaching the boundary of the maximal Cauchy development?

\section{Generalized Kasner Variables and Velocity-Dominated
Approximation}

Let $K_i^{\;j}$ and $g_{ij}$ be the extrinsic curvature and spatial
metric tensors induced  by an arbitrary foliation $\Sigma_t$  on the
spacetime, and let $\tau=Tr(K)$ and $\hat {K}_i^{\;j}=K_i^{\;j}/
\tau$. There exists a {\em unique} triad $E_{(p)}^{\;i}$ of spacelike
vector fields which puts $K_i^{\;j}$ and $g_{ij}$ into the following
form:
\begin{eqnarray}
E\,g\,E^T &=&\left(
\begin{array}{ccc}
\; \tau^{-2p_1}& 0 & 0 \\
\; 0&  \tau^{-2p_2}& 0 \\
\; 0& 0 & \tau^{-2p_3}
\end{array}
              \right),\\
&\;& \nonumber\\
E\,\hat K\, E^{-1}&=&\left(
\begin{array}{ccc}
\; p_1& 0 & 0 \\
\; 0&  p_2& 0 \\
\; 0& 0 & p_3
\end{array}
              \right),
\end{eqnarray}
provided $p_1\neq p_2\neq p_3$ and $\tau >0$. In that case  there is a
one to one mapping between the old dynamical variables $(K,g)$ and the
new ones $(E,\tau,p)$ with $\Sigma \;p_q=1$.    If we restrict
ourselves to a subclass of cosmological spacetimes that admit
constant-mean-curvature (CMC) foliation,  which is unique if it exists
\cite{marsden-tipler80}, we obtain what I call generalized Kasner
variables (GKV): a triad of spacelike vector fields $E_{(q)}^{\;i}$
and three Kasner exponents $p_q$ (two of them independent),
invariantly defined throughout the spacetime. Since the
globally-hyperbolic spacetimes with surfaces of infinite
mean-curvature (crushing singularities) have Cauchy boundaries exactly
at those surfaces, we can analyse the Cauchy boundaries by taking limit
$\tau\rightarrow \pm \infty$. In spacetimes with no crushing
singularities, however, there is no generic way to lock onto the
Cauchy boundary, whose elusiveness prevents us from analysing the
geometry near it. So we should restrict our attention to the crushing
singularities.

In order to study the behavior of metric near the spacetime's Cauchy
boundary, I use  a  perturbative VDA method previously developed by
Moncrief and myself. It was applied successfully to $T^3\times R$ Gowdy
spacetimes, where it was found, by an inductive argument, that the
VDA is valid to an arbitrarily high perturbative order
\cite{grubisic-moncrief}. The method is applicable to spacetimes with
no isometries as well, and, in a nutshell, consists of modifying
Einstein's equations by multiplying the shift vector in the ADM action
by an $\epsilon$,
\begin{equation}
I_{ADM} \rightarrow I_{ADM\epsilon}=
\int \left[ N H + \epsilon N^i H_i \right],
\end{equation}
and then expanding  the solution of the modified Einstein's equations
with respect to that $\epsilon$, about $\epsilon=0$. The introduction
of $\epsilon$ is just a device to obtain a well defined  series of
recursively solvable perturbative equations which start with the
generalized Kasner solution as the zeroth-order term. The case we are
interested in is $\epsilon=1$.

After rewriting Einstein's equations in terms of the GKV, and
choosing the time $t$ to be some monotonous function of the mean
curvature $\tau$, we find that in the zeroth order the GKV are time
independent, that the lapse function $N$ depends only on time $t$, and
that the Hamiltonian constraint gives an additional restriction on
(generalized) Kasner exponents, which together with the initial trace
constraint gives just the standard Kasner relations for the zeroth
order functions (only of ``spacelike coordinates''):
\begin{equation}
\Sigma\; p_q^2= \Sigma \;p_q =1
\end{equation}
The zeroth order solution is just the generalized Kasner solution.

All higher order terms, in turn, are uniquely determined by the
initial data for the zeroth order solution and, using an inductive
argument,  it can be showed that,  as $\tau\rightarrow \infty$, they
become negligible  when compared with the zeroth order provided
\begin{equation} \vec E_1\; (\vec \nabla \times \vec E_1)=0,
\label{condition} \end{equation} where $\vec E_1$ is the eigenvector
with the only possible negative Kasner exponent. Roughly, the rate of
decay is faster the higher the order.

\section{Conclusions} The condition (\ref{condition}) has already been
established by BKL \cite{bkl82} as a sufficient condition for the
decay of the first-order correction. Using the generalized Kasner
variables I showed that (\ref{condition}) is also sufficient condition
for the decay of all higher orders in the expansion, asymptotically as
$\tau\rightarrow \infty$, and, therefore, for the VDA to be valid
perturbatively. Even though I did not use a  synchronous foliation,
the fact that the lapse function $N$ asymptotically approaches a
spatially homogeneous function indicates that the CMC foliation
asymptotically becomes a synchronous foliation when approaching $\tau=
\infty$.

\end{document}